\documentclass[floatfix,aps,pra,twocolumn,a4paper,superscriptaddress,nofootinbib,balancelastpage,reprint]{revtex4-1}
\usepackage[hidelinks,breaklinks=true]{hyperref}
\usepackage[english]{babel}
\usepackage{amsopn,amsthm,dsfont}
\DeclareMathOperator{\Tr}{Tr}
\newcommand{\id}{\mathds{1}}
\usepackage{mathtools}
\mathtoolsset{showonlyrefs=true}
\usepackage{tikz}
\usetikzlibrary{positioning}
\begin{document}
\title{Maximal incompatibility of locally classical behavior and global causal order\\in multi-party scenarios}
\author{\"Amin Baumeler}
\affiliation{Faculty of Informatics, Universit\`{a} della Svizzera italiana, Via G. Buffi 13, 6900 Lugano, Switzerland}
\author{Adrien Feix}
\affiliation{Faculty of Physics, University of Vienna, Boltzmanngasse 5, 1090 Vienna, Austria}
\affiliation{Institute for Quantum Optics and Quantum Information (IQOQI), Austrian Academy of Sciences, Boltzmanngasse 3, 1090 Vienna, Austria}
\author{Stefan Wolf}
\affiliation{Faculty of Informatics, Universit\`{a} della Svizzera italiana, Via G. Buffi 13, 6900 Lugano, Switzerland}
\begin{abstract}
	\noindent
Quantum theory in a global space-time gives rise to non-local correlations, which cannot be explained causally in a satisfactory way; this motivates the study of theories with reduced global assumptions.
Oreshkov, Costa, and Brukner (2012) proposed a framework in which quantum theory is valid locally but where, at the same time, no global space-time, {\em i.e.}, predefined causal order, is assumed beyond the absence of logical paradoxes. 
It was shown for the two-party case, however, that a global causal order always emerges in the classical limit. 
Quite naturally, it has been conjectured that the same also holds in the multi-party setting.
We show that counter to this belief, classical correlations locally compatible with classical probability theory exist that allow for deterministic signaling between three or more parties incompatible with any predefined causal order.
\end{abstract}

\maketitle

\section{Motivation and main result}
According to Bell~\cite{Bell:1981gx}, correlations cry out for explanation.
In such a spirit, already Einstein, Podolsky, and Rosen~(EPR)~\cite{Einstein:1935rr} had asked for an extension of quantum theory that incorporates a {\em causal explanation\/}~\cite{Reichenbach:1956vl,Wiseman:2014vo} of the correlations arising when two parts of an entangled quantum state are measured.
Such an explanation can describe the emergence of the correlations either through {\em pre-shared information\/} or through {\em influences}.
Because of relativity, EPR argued further, the latter cannot be the cause of such correlations.
Later, (finite-speed) influences were ruled out by theory~\cite{Wood:2012vw,Barnea:2013ev} and experiments~\cite{Aspect:1982ja,Suarez:1997ds,Weihs:1998cc,Aspect:1999gs,Rowe:2001ic,Salart:2008ku,Giustina:2013js}.
Therefore, still according to EPR's reasoning, physical quantities need to be {\em predefined}.
This, however, had been rejected by Bell~\cite{Bell:1964ws} under the assumption that spatially separated settings can be chosen (at least partially~\mbox{\cite{Hall:2011je,Barrett:2011cz,Putz:2014vk}}) freely and independently; such correlations are called {\em non-local}.
Remarkably, this means that~there are {\em not predefined yet correlated\/} physical quantities emerging in a space-like separated way.
However, although the EPR program as such may have failed, it seems natural to continue to ask for a causal explanation of non-local correlations.
A possible approach is to refrain from considering space-time as fundamental, treating it as emerging (potentially along with other macroscopic quantities) from a deeper fundament~\mbox{\cite{Parmenides,Wootters:1984gt,Bombelli:1987ez,MauroDAriano:2011in}}~---~comparably to temperature.
A step in this direction was taken by Hardy~\cite{Hardy:2005wj,Hardy:2007bk} with his program of merging general relativity with quantum theory, in which he proposes to extend the latter to {\em dynamical causal orders}, a feature of relativity (see~\cite{Brukner:2014if} for a recent review on quantum theory and causality).
Chiribella, D'Ariano, and Perinotti~\cite{Chiribella:2009bh,Chiribella:2013bk} studied quantum supermaps called ``quantum combs'' that allow for superpositions of causal orders.
Based on Hardy's idea, Oreshkov, Costa, and Brukner~\cite{Oreshkov:2012uh} developed a framework for quantum correlations without predefined causal order~by dropping the assumption of a {\em global background time\/} while keeping the~assumptions that {\em locally}, nature is described by quantum theory and that {\em no logical paradoxes\/} arise.
Some causal structures emerging from this framework cannot be predefined~\cite{Oreshkov:2012uh,Baumeler:2013wy,Brukner:2014vo} --- just like physical quantities exhibiting non-local correlations~\mbox{\cite{Bell:1964ws,Cirelson:1980fp,Greenberger:1989vx,Greenberger:1990it}}. 
If, in the {\em two-party\/} case, we consider the classical limit of the quantum systems, {\em i.e.}, enforce both parties' physics to be described by {\em classical probability theory\/} (instead of quantum theory), then a predefined causal order \emph{always emerges}~\cite{Oreshkov:2012uh}.
This is in accordance with our experience and, hence, natural and unsurprising; it strongly indicates that the same may hold in the {\em multi-party\/} case~\cite{Costa:2013vc}.
This, however, fails to be true, as we show in the present work.

\section{Input-output systems, and\\causal order}
By definition, measurement settings and outcomes are classical, {\em i.e.}, perfectly distinguishable.
Therefore, we think of physical systems as black boxes which we probe with classical inputs and that respond with classical outputs.
When taking this perspective, we describe all \emph{physical quantities}, {\em i.e.}, {\em outputs}, as functions of \emph{inputs}. The~party~$S$ is described by a set of \emph{inputs\/}~$V(S)=\{A_i\}_I$ chosen \emph{freely\/} by~$S$, and by a set of \emph{outputs\/}~\mbox{$Q(S)=\{X_j\}_J$}~the~party can access (instantiations of the inputs and outputs are denoted by the same letters but in lowercase).
Since we refrain from assuming global {\em space-time\/} as given {\em a priori},  we cannot define {\em free randomness\/} based on such a causal structure, as done elsewhere~\cite{Colbeck:2011hw,Ghirardi:2013bb}.
Instead, we take the concept of {\em free~randomness\/} as fundamental~--- in accordance with a recent trend to derive properties of quantum theory from information-theoretic principles~\cite{Caves:2002hf,Brukner:2003,Clifton:2003di,Brassard:2005di,Pawiowski:2009dt,Chiribella:2011jb,Muller:2013gn,Pfister:2013ik} --- and {\em postulate\/} inputs as being \emph{free}.

Outputs are functions of inputs.
Based on this relationship, we define {\em causal order}.
If an output~$X_j$ is a function of~$A_i$, we say that~$X_j$ \emph{causally depends\/} on~$A_i$ and is in the \emph{causal future\/} of~$A_i$ or, equivalently, that~$A_i$ is in the {\em causal past\/} of~$X_j$, denoted by~$X_j \succeq A_i$ or~$A_i \preceq X_j$.
The negations of these relations are denoted by~$\not\succeq$ and~$\not\preceq$.
This definition does neither induce a causal order between outputs nor between inputs nor between any output and the input it does not depend on.

\begin{figure}
	\centering
	\includegraphics[scale=1]{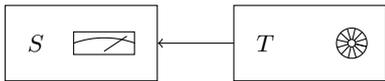}
%	\beginpgfgraphicnamed{pra-f1}
%	\begin{tikzpicture}
%		\node[draw,shape=rectangle,minimum width=2cm,minimum height=1cm] (S) {};
%		\node[right=0.3cm of S.west,inner sep=0pt,outer sep=0pt] (St) {$S$};
%		\node[left=0.3cm of S.east,draw,shape=rectangle,minimum width=0.8cm,minimum height=0.3cm] (M) {};
%		\draw (M.east) arc (70:110:1.18);
%		\draw[-] (M.south)++(0cm,0.05cm) -- ++(0.3cm,0.2cm);
%		\node[draw,shape=rectangle,minimum width=2cm,minimum height=1cm,right=of S,inner sep=0pt,outer sep=0pt] (T) {};
%		\node[right=0.3cm of T.west,inner sep=0pt,outer sep=0pt] (Tt) {$T$};
%		\node[left=0.4cm of T.east,draw,shape=circle,inner sep=0pt,outer sep=0pt,minimum size=0.1cm] (K) {};
%		\node[draw,shape=circle,inner sep=0pt,outer sep=0pt,minimum size=0.4cm] (K2) at (K) {};
%		\foreach \angle in {
%			-13+0*30,
%			-13+1*30,
%			-13+2*30,
%			-13+3*30,
%			-13+4*30,
%			-13+5*30,
%			-13+6*30,
%			-13+7*30,
%			-13+8*30,
%			-13+9*30,
%			-13+10*30,
%			-13+11*30,
%			-13+12*30
%		}
%		{
%			\draw (K.center)++(\angle:0.05cm) -- +(\angle:0.15cm);
%		}
%		\draw[->] (T.west) -- (S.east);
%	\end{tikzpicture}
%	\endpgfgraphicnamed
	\caption{If party~$T$ can freely choose an input (here, visualized by a knob), and party~$S$ can read off an output that depends on~$T$'s input, then~$T$ can {\em signal\/} to~$S$, which implies that~$S$ is in the {\em causal future\/} of~$T$ ($S\succeq T$).}
	\label{fig:signaling}
\end{figure}
Let us introduce a second party~$T$ described by the set of inputs~$V(T)=\{B_k\}_K$ and with access to the outputs~\mbox{$Q(T)=\{Y_\ell\}_L$}.
Outputs can depend on inputs of both parties.
If party~$S$ has an output that depends on an inputs of~$T$, then we say that~$T$ can \emph{signal\/} to~$S$ (see Figure~\ref{fig:signaling}).
In the following, we will assume {\em unidirectional\/} signaling: If~$S$ can signal to~$T$, then~$T$ cannot signal to~$S$.
This enables us to causally order parties.
If at least one output of~$S$ depends on an input of~$T$,
but no output of~$T$ depends on any input of~$S$ (which is the condition for unidirectional signaling),
then~$S$ is in the causal future of~$T$.
Formally, if there exist~$X\in Q(S)$ and~$B\in V(T)$ fulfilling~$X\succeq B$ and if for all~$Y\in Q(T)$ and for all~$A\in V(S)$, the relation~$Y\not\succeq A$ holds, then we have~$S\succeq T$.

Consider a two-party scenario with parties~$S,T$, each having a single input~$A,B$, a single output~$X,Y$, respectively, and a shared random variable~$\Lambda$.
We call a theory compatible with \emph{predefined causal order\/} if all achievable probability distributions~$P(x,y|a,b)$ can be written as a convex mixture of possible causal orders, {\em i.e.},
\begin{align}
	P(x,y|a,b)&=\notag\\
	\Pr(\alpha)&\sum_\lambda \Pr(\lambda|\alpha)\Pr(x|a,\lambda,\alpha)\Pr(y|a,b,\lambda,\alpha)\notag\\
	+\,\Pr(\neg\alpha)&\sum_\lambda \Pr(\lambda|\neg\alpha)\Pr(x|a,b,\lambda,\neg\alpha)\Pr(y|b,\lambda,\neg\alpha)\notag
	\,,
\end{align}
where~$\alpha$ is the event~$S\preceq T$, and~$\lambda$ is an instantiation of~$\Lambda$ that depends on a input not in either of the sets~$V(S)$ or~$V(T)$.
For more than two parties, the definition of predefined causal order becomes more subtle.
Suppose we have three parties~$S$,~$T$, and~$U$, where~$S$ is in the causal past of both~$T$ and~$U$.
We call a causal order \emph{predefined\/} even if~$S$ is free to choose the causal order between~$T$ and~$U$~\cite{Fabio,Christina}.
In general, in a predefined causal order, a party is allowed to determine the causal order between all parties in her causal future.
Hence, a theory with the parties~$S_0,\dots,S_{n-1}$, inputs~$A_0,\dots,A_{n-1}$~(shorthand~$\vec A$), and outputs~$X_0,\dots,X_{n-1}$ ($\vec X$), respectively, is compatible with {\em predefined causal order\/} if all achievable probability distributions~$P(\vec x|\vec a)$ can be written as
\begin{align}
	P(\vec x|\vec a)&=\notag\\
	\sum_{i=0}^{n-1}&\Pr\left(\alpha_i\wedge\neg\alpha_0\wedge\dots\wedge\neg\alpha_{i-1}\right)\Pr(\vec x|\vec a,S_i\text{ is first})
	\notag
	\,,
\end{align}
where~$\alpha_i$ is the event that each party~$S_{j(\not=i)}$ either is in the causal future of~$S_i$~($S_i\preceq S_j$) or has no causal relation with~$S_i$ ($S_i \not\preceq S_j$ and~\mbox{$S_i \not\succeq S_j$}).
The~term~$\Pr(\vec x|\vec a,S_i\text{ is first})$ is a convex mixture of distributions compatible with the causal structures in which~$S_i$ is first and chooses the causal order between the remaining parties.

\section{Game}
The following multi-party game cannot be won in a scenario with predefined causal order.
Denote by~$S_0,\dots,S_{n-1}$ the parties that participate in the game.
Each party~$S_i$ has a uniformly distributed binary input~$A_i$ as well as a binary output~$X_i$ and access to the shared random variable~$M$ uniformly distributed in the range~$\{0,\dots,n-1\}$.
The random variable~$M$ belongs to a dummy party (we need her as a source of shared randomness).
For given~$M=m$, the game is won whenever~$S_m$'s output~$X_m$ is the parity of the inputs to all other parties,~{\em i.e.},~$X_m=\bigoplus_{i\not=m}A_i$.
Therefore, the success probability for winning the game is
\begin{align}
	\label{eq:succprobab}
	p_\text{succ}=\frac{1}{n}\sum_{m=0}^{n-1}\Pr\left(X_m=\bigoplus_{i\not=m}A_i\,\middle|\,M=m\right)
	\,.
\end{align}

In a setup with predefined causal order, this success probability is upper bounded by~$1-1/(2n)$.
To see this, note that if, without loss of generality,~$S_0$ is first, she will remain first.
For~$n>2$, the last party can be specified by~$S_0$.
Thus, all the terms of the sum in expression~\eqref{eq:succprobab} are~$1$ except for the first summand, reflecting the fact that~$S_0$ herself has to guess the parity of the other's inputs, which is~$1/2$.
By repeating the experiment~$\omega(n)$ times, one can bring the winning probability arbitrarily closely to~$0$.

\section{Framework for classical correlations without causal order}
Instead of assuming that locally, nature is described by {\em quantum theory\/}~\cite{Oreshkov:2012uh}, we take the {\em classical limit\/} of the systems and thus assume that locally, nature is described by \emph{classical probability theory}.
In addition to this, we require the probabilities of the outcomes to be non-negative and to sum up to~$1$; this excludes logical paradoxes~\cite{Oreshkov:2012uh,Chiribella:2013bk}.
We suppose that each party has a closed laboratory that can be opened once --- which is when the only interaction with the environment happens.
When a laboratory is opened, the party receives, manipulates, and outputs a state.
Thus, in the setting of local validity of classical probability theory, such a laboratory is described by a conditional probability distribution~$P_{O|I}$, where~$I$ is the input to and~$O$ is the output from the laboratory.

\begin{figure}
	\centering
	\includegraphics[scale=1]{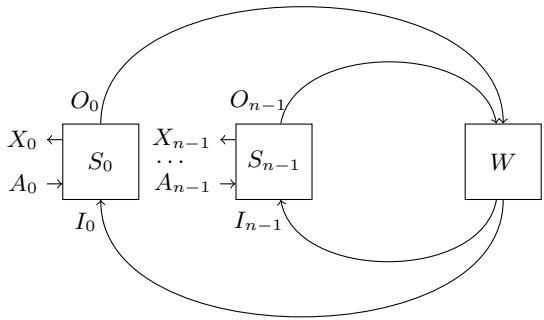}
%	\beginpgfgraphicnamed{pra-f2}
%	\begin{tikzpicture}
%		\node[draw,rectangle,minimum width=0.5cm,minimum height=1cm,minimum width=1cm] (S) {$S_0$};
%		\draw[->] (S.150) -- ++(-0.2,0) node [left] {$X_0$};
%		\draw[<-] (S.210) -- ++(-0.2,0) node [left] {$A_0$};
%		\node[right=0.1cm of S] (D) {\dots};
%		\node[draw,rectangle,minimum width=0.5cm,minimum height=1cm,minimum width=1cm,right=0.5cm of D] (T) {$S_{n-1}$};
%		\draw[->] (T.150) -- ++(-0.2,0) node [left] {$X_{n-1}$};
%		\draw[<-] (T.210) -- ++(-0.2,0) node [left] {$A_{n-1}$};
%		\node[draw,rectangle,minimum width=1cm,minimum height=1cm,right=2cm of T] (W) {$W$};
%		\draw[->] (S) to[out=90,in=90] (W);
%		\draw[<-] (S) to[out=270,in=270] (W);
%		\draw[->] (T) to[out=80,in=100] (W);
%		\draw[<-] (T) to[out=280,in=260] (W);
%		\node (SO) at ([shift={(-0.2cm,0.8cm)}]S) {$O_0$};
%		\node (SI) at ([shift={(-0.2cm,-0.8cm)}]S) {$I_0$};
%		\node (TO) at ([shift={(-0.2cm,0.8cm)}]T) {$O_{n-1}$};
%		\node (TI) at ([shift={(-0.2cm,-0.8cm)}]T) {$I_{n-1}$};
%	\end{tikzpicture}
%	\endpgfgraphicnamed
	\caption{Party~$S_i$ is described by~$P_{X_i,O_i|A_i,I_i}$. Her output~$O_i$ is fed to~$W$, which describes {\em everything\/} outside the laboratories. Therefore,~$W$ also provides the input~$I_i$ and is described by~$W=P_{I_0,\dots,I_{n-1}|O_0,\dots,O_{n-1}}$.}
	\label{fig:connections}
\end{figure}
Let us consider the parties as described in the game.
We denote the input to~$S_i$ by~$I_i$ and the output from~$S_i$ by~$O_i$.
Therefore, the~$i$th local laboratory is described~by the distribution~$P_{X_i,O_i|A_i,I_i}$.
As we do not make global assumptions other than that the overall picture should describe a probability distribution, we describe everything outside the laboratories by the distribution~$W$~(see~Figure~\ref{fig:connections}) satisfying the condition that for any choice of~$a_0,\dots,a_{n-1}$, {\em i.e.}, for any probability distribution~$P_{X_0,O_0|I_0},\dots,P_{X_{n-1},O_{n-1}|I_{n-1}}$, the values of the product with~$W$, {\em i.e.}, the values of
\begin{align}
	P_{X_0,O_0|I_0}\cdot\ldots\cdot P_{X_{n-1},O_{n-1}|I_{n-1}}\cdot P_{I_0,\dots,I_{n-1}|O_0,\dots,O_{n-1}}
	\,,
	\end{align}
and in general of
\begin{align}
	P_{O_0|I_0}\cdot\ldots\cdot P_{O_{n-1}|I_{n-1}}\cdot P_{I_0,\dots,I_{n-1}|O_0,\dots,O_{n-1}}
	\,,
\end{align}
are non-negative and sum up to~$1$.
For tackling this condition formally, we represent a probability distribution~$P_X$ as a real positive diagonal matrix~$\hat P_X$ with trace~$1$ and diagonal entries~$P_X(x)$.
A conditional probability distribution~$P_{X|Y}$ is a collection of (unconditional) probability distributions~$P_{X|Y=y}$ for each value of~$y$.
Thus, we represent~$P_{X|Y}$ similarly, yet with trace~$|{\cal Y}|$, where~${\cal Y}$ is the set of values~$y$ can take, and we use the symbol~$\hat P_{X|Y}$.
The values~$P_{X|Y=y}(x)$ are ordered with respect to the ordering of the subscripts of~$P_{X|Y}$, {\em e.g.}, for binary~$X$ and~$Y$ the matrix~$\hat P_{X|Y}$ is
\begin{align}
	\begin{pmatrix}
		P_{X|Y=0}(0) & 0 & 0 & 0\\
		0 & P_{X|Y=1}(0) & 0 & 0\\
		0 & 0 & P_{X|Y=0}(1) & 0\\
		0 & 0 & 0 & P_{X|Y=1}(1)
	\end{pmatrix}
	\,.
\end{align}
The condition that the probabilities~$P_{X|Y=y}(x)$ sum up to~$1$ for fixed~$y$ is reflected by the condition that if we trace out~$X$ from the matrix~$\hat P_{X|Y}$ (denoted by~$\Tr_X \hat P_{X|Y}$), we are left with the identity.
The product of two distributions~$\hat P_X$ and~$\hat P_Y$ in the matrix representation corresponds to the tensor product denoted by~$\hat P_X\otimes\hat P_Y$.
To obtain the marginal distribution from a joint distribution, we use the partial trace.
This implies that the output state of a laboratory~$P_{O_i|I_i}$, given the input state~$P_{I_i}$, is~$\Tr_{I_i}(\hat P_{O_i|I_i}\cdot(\id_{O_i}\otimes\hat P_{I_i}))$, where~$\id_{O_i}$ is the identity matrix with the same dimension as~$\hat P_{O_i}$.
This allows us to use the framework of Oreshkov, Costa, and Brukner~\cite{Oreshkov:2012uh}, where we restrict ourselves to diagonal matrices, {\em i.e.}, all objects ($W$ and local operations) are \mbox{simultaneously} diagonalizable in the computational basis and can, hence, be expressed using the identity~$\id$ and the Pauli matrix~$\sigma_z$.
We know from their framework~\cite{Oreshkov:2012uh} that if we express~$P_{I_0,\dots,I_{n-1}|O_0,\dots,O_{n-1}}$ as a matrix~$W=c\sum_i g_i$, where~$c$ is a normalization constant and~$g_i=R_{i,0}\otimes\dots\otimes R_{i,n-1}\otimes T_{i,0}\otimes\dots\otimes T_{i,n-1}$.
For every~$i$, the summand~$g_i$ represents a channel from all~$S_j$ with~$\Tr T_{i,j}=0$ to all~$S_k$ with~$\Tr R_{i,k}=0$.
In order to avoid logical paradoxes,~$g_i$ must describe a channel where at least one party is a recipient without being a sender~\cite{Oreshkov:2012uh}. 
In other words,~$g_i$ must either be the identity or there exists~$j$ such that~$T_{i,j}=\id$ and~$\Tr R_{i,j}=0$.

\section{Winning the game perfectly}
To win the game using this framework, we need to provide the distribution~$P_{I_0,\dots,I_{n-1}|O_0,\dots,O_{n-1}}$ and all distributions describing the laboratories.
For that purpose, we use the fact that if a set~$\{g_i\}_I$ of matrices with all eigenvalues in~$\{-1,1\}$ forms an Abelian group with respect to matrix multiplication, then~$\sum_{i\in I} g_i$ is a positive semi-definite matrix.
To prove this, take the eigenvector~$\vec v$ which has the smallest eigenvalue~$\lambda_\text{min}$, {\em i.e.},
\begin{align}
	\sum_{i\in I} g_i\vec v=\sum_{i\in I} \lambda^{\vec v}_i\vec v=\lambda_\text{min}\vec v
	\,,
\end{align}
where~$\lambda^{\vec v}_i$ is the eigenvalue of~$g_i$ with respect to the eigenvector~$\vec v$.
Let~$g_{i_0}$ be an element contributing negatively to~$\lambda_\text{min}$, {\em i.e.},~$g_{i_0}\vec v=-\vec v$.
As the set forms a group, for every~$j$ there exists a~$k\not=j$ such that~$g_{i_0}\cdot g_j=g_k$.
This implies~$-\lambda^{\vec v}_j=\lambda^{\vec v}_k$ and~$\sum_{i\in I}\lambda^{\vec v}_i=0$.

\subsection{Construction of~$W_n$ for odd~$n$}
We construct the distribution~$P_{I_0,\dots,I_{n-1}|O_0,\dots,O_{n-1}}$ for odd~\mbox{$n>2$}.
Let~$\{g_i\}_I$ be the set of matrices that can be written as~\mbox{$g_i=g_{i,1}\otimes g_{i,2}\otimes\dots\otimes g_{i,n}$}, with the objects~\mbox{$g_{i,k}\in\{\id,\sigma_z\}$},
and with an even number of~$\sigma_z$'s for each~$i\in I$.
We use the notation~$g_{i,j:k}$ to denote the matrix~$g_{i,j}\otimes g_{i,j+1}\otimes\dots\otimes g_{j,k}$.
The fact~$\sigma_z^2=\id$ implies that the product~$g_i\cdot g_j$, for every~$i,j\in I$, is a tensor product of~$\id$ and~$\sigma_z$ with an even number of~$\sigma_z$'s, and is thus an \mbox{element} of~$\{g_i\}_I$.
Furthermore, all elements mutually commute, have all eigenvalues in~$\{-1,1\}$ and, hence, each element is an involution.
Therefore, their sum is a positive semi-definite matrix.
The distribution~$P_{I_0,\dots,I_{n-1}|O_0,\dots,O_{n-1}}$ as a matrix~$W_n$ is built by taking the sum over all group elements, where the matrix~$g_{i,k}$ of the group element~$g_i$ contributes to the input~$I_k$ of party~$S_k$, and to the output~$O_{k-1\bmod n}$ of the party labeled by~\mbox{$(k-1\bmod n)$},
\begin{align}
	W_n=\hat P_{I_0,\dots,I_{n-1}|O_0,\dots,O_{n-1}}=\frac{1}{2^n}\sum_{i\in I} g_i\otimes g_{i,2:n}\otimes g_{i,1}
\,.
\end{align}
By construction,~$W_n$ is positive semi-definite, {\em i.e.}, all probabilities are positive.
Because~$n$ is odd, there exists for each group element $g_i$ ($\not= \id$) at least one position~$k$ such that~$g_{i,k}\otimes g_{i,k+1\bmod n}=\sigma_z\otimes\id$, which excludes logical paradoxes.
Furthermore, for every~$i\in \{0,1,\dots,n-1\}$ the object~$W_n$ contains the channel from all parties~$S_{j(\not= i)}$ to~$S_i$ --- a condition to perfectly win the game.

\subsection{Example:~$W_3$}
For illustration, we construct~$W_3$.
The group from which~$W_3$ is constructed is~$\{g_0,g_1,g_2,g_3\}$ with the group elements
\begin{align}
	g_0&=\id\otimes\id\otimes\id\,,\\
	g_1&=\id\otimes\sigma_z\otimes\sigma_z\,,\\
	g_2&=\sigma_z\otimes\id\otimes\sigma_z\,,\\
	g_3&=\sigma_z\otimes\sigma_z\otimes\id\,.
\end{align}
The matrix~$W_3$ is thus
\begin{align}
	W_3=\frac{1}{8}
	\bigl(\id^{\otimes 6}&+\id\otimes\sigma_z\otimes\sigma_z\otimes\sigma_z\otimes\sigma_z\otimes\id\\
	&+\sigma_z\otimes\id\otimes\sigma_z\otimes\id\otimes\sigma_z\otimes\sigma_z\\
	&+\sigma_z\otimes\sigma_z\otimes\id\otimes\sigma_z\otimes\id\otimes\sigma_z\bigr)
	\,.
\end{align}
The second summand of~$W_3$ represents a channel from~$S_0,S_1$ to~$S_2$, the third summand represents a channel from~$S_1,S_2$ to~$S_0$, and finally, the last summand represents a channel from~$S_0,S_2$ to~$S_1$.

\begin{figure}
	\centering
	\includegraphics[scale=1]{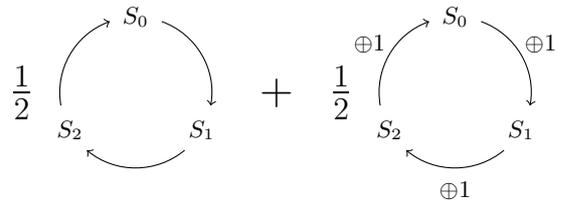}
%	\beginpgfgraphicnamed{pra-f3}
%	\begin{tikzpicture}
%		\def\r{1}
%		\def\d{20}
%		\def\dx{2.1}
%		\def\hd{.5}
%
%		\draw[->] (-\dx,0)++(90-\d:\r) arc (90-\d:-30+\d:\r);
%		\draw[->] (-\dx,0)++(-30-\d:\r) arc (-30-\d:-150+\d:\r);
%		\draw[->] (-\dx,0)++(-150-\d:\r) arc (-150-\d:-270+\d:\r);
%		\draw (-\dx,0)++(90:\r) node (A) {$S_0$};
%		\draw (-\dx,0)++(-30:\r) node (B) {$S_1$};
%		\draw (-\dx,0)++(-150:\r) node (C) {$S_2$};
%
%		\draw[->] (\dx,0)++(90-\d:\r) arc (90-\d:-30+\d:\r);
%		\draw[->] (\dx,0)++(-30-\d:\r) arc (-30-\d:-150+\d:\r);
%		\draw[->] (\dx,0)++(-150-\d:\r) arc (-150-\d:-270+\d:\r);
%		\draw (\dx,0)++(90:\r) node (A2) {$S_0$};
%		\draw (\dx,0)++(-30:\r) node (B2) {$S_1$};
%		\draw (\dx,0)++(-150:\r) node (C2) {$S_2$};
%		\draw (\dx,0)++(90-60:\r+.3) node (n1) {$\oplus 1$};
%		\draw (\dx,0)++(-30-60:\r+.3) node (n2) {$\oplus 1$};
%		\draw (\dx,0)++(-150-60:\r+.3) node (n3) {$\oplus 1$};
%
%		\draw (-\hd/2,0) node (PLUS) {\LARGE $+$};
%		\draw (-\dx-\r-\hd,0) node (half1) {\LARGE $\frac{1}{2}$};
%		\draw (\dx-\r-\hd,0) node (half2) {\LARGE $\frac{1}{2}$};
%	\end{tikzpicture}
%	\endpgfgraphicnamed
	\caption{The conditional probability distribution~$W_3$ is a mixture of a circular identity channel with a circular bit-flip channel.}
	\label{fig:w3loops}
\end{figure}

It can easily be verified that if the three parties~$S_0$,~$S_1$, and~$S_2$ input~\mbox{$P_{O_0,O_1,O_2}(o_0,o_1,o_2)=1$} into~$W_3$, then~$W_3$ outputs the distribution
\begin{align}
	P_{I_0,I_1,I_2}(i_0,i_1,i_2)=\begin{cases}
		\frac{1}{2},&\text{$i_0=o_2$,~$i_1=o_0$,~$i_2=o_1$,}\\
		\frac{1}{2},&\text{$i_0=\bar o_2$,~$i_1=\bar o_0$, $i_2=\bar o_1$,}\\
		0, &\text{otherwise,}
	\end{cases}
\end{align}
where~$\bar o_i=o_i\oplus 1$.
Therefore,~$W_3$ implements a uniform mixture of the loops where the input of party~\mbox{$S_{i\bmod 3}$} is sent to party~\mbox{$S_{i+1\bmod 3}$}, and where the input of party~\mbox{$S_{i\bmod 3}$} is flipped and sent to~\mbox{$S_{i+1\bmod 3}$} (see Figure~\ref{fig:w3loops})~\cite{Caslav}.
It is evident from Figure~\ref{fig:w3loops} that logical paradoxes are not possible.
If all intermediate parties forward what they receive (by applying any reversible transformation), both loops (see Figure~\ref{fig:w3loops}) cancel each other out, {\em i.e.}, the correlations interfere destructively.
Then again, if one intermediate party does not forward what she receives, the loop is broken.
Conversely, any party can signal to her predecessor in the loop, because then an even number of bit-flips are applied, and thus the correlations interfere constructively.
The same reasoning holds for any~$W_n$ for odd~$n>2$.

\subsection{Construction of~$W_n$ for even~$n$}
The above construction works for odd~$n>2$.
For even~$n$, the group contains the element~$\sigma_z^{\otimes n}$, which leads to a logical paradox since all inputs are correlated to all outputs~\cite{Oreshkov:2012uh}.
This can also be seen in Figure~\ref{fig:w3loops}, where for even~$n$, the sum of both channels leads to a logical paradox.
In this case ($n$ even), we double the dimensions~of the output of the second-to-last party and of the input of the last party, and construct the distribution based on the group of matrices for the case of~$n-1$.
Let~$\{g_i\}_I$ be the group used to construct~$W_{n-1}$.
The set for~$n$ parties is the Abelian subgroup~$\{g_i \otimes g'_i\}_I\cup\{\bar g_i\otimes g'_i\}_I$, where~\mbox{$g'_i=g_{i,1}\otimes g_{i,2}$} and~$\bar g_i=g_i \cdot \sigma_z^{\otimes (n-1)}$.

The distribution~$P_{I_0,\dots,I_{n-1}|O_0,\dots,O_{n-1}}$ as a matrix~$W_n$ is constructed as before, with the exception that~$g'_i$ is considered a {\em single\/} submatrix, 
\begin{align}
	W_n=\frac{1}{2^{n+1}}\sum_{i\in I}&\Bigl(
	g_i\otimes g'_i \otimes g_{i,2:n-1}\otimes g'_i\otimes g_{i,1}\notag\\
	&+\bar g_i\otimes g'_i \otimes \bar g_{i,2:n-1}\otimes g'_i\otimes \bar g_{i,1}
	\Bigr)
\,.
\end{align}
Again, by construction,~$W_n$ fulfills all requirements and contains all channels required to perfectly win the game.

\subsection{Example:~$W_4$}
As an example, we construct the matrix~$W_4$. The group~$\{h_0,h_1,h_2,h_3,h_4,h_5,h_6,h_7\}$ for~$W_4$ is constructed from the group~$\{g_0,g_1,g_2,g_3\}$ and has the elements
\begin{align}
	h_0&=g_0\otimes g'_0=(\id\otimes\id\otimes\id)\otimes(\id\otimes\id)\,,\\
	h_1&=g_1\otimes g'_1=(\id\otimes\sigma_z\otimes\sigma_z)\otimes(\id\otimes\sigma_z)\,,\\
	h_2&=g_2\otimes g'_2=(\sigma_z\otimes\id\otimes\sigma_z)\otimes(\sigma_z\otimes\id)\,,\\
	h_3&=g_3\otimes g'_3=(\sigma_z\otimes\sigma_z\otimes\id)\otimes(\sigma_z\otimes\sigma_z)\,,\\
	h_4&=\bar g_0\otimes g'_0=(\sigma_z\otimes\sigma_z\otimes\sigma_z)\otimes(\id\otimes\id)\,,\\
	h_5&=\bar g_1\otimes g'_1=(\sigma_z\otimes\id\otimes\id)\otimes(\id\otimes\sigma_z)\,,\\
	h_6&=\bar g_2\otimes g'_2=(\id\otimes\sigma_z\otimes\id)\otimes(\sigma_z\otimes\id)\,,\\
	h_7&=\bar g_3\otimes g'_3=(\id\otimes\id\otimes\sigma_z)\otimes(\sigma_z\otimes\sigma_z)
	\,.
\end{align}
The matrix~$W_4$ is thus
\begin{align}
	W_4&=\frac{1}{32}
	\bigr(\id^{\otimes 10}\\
&+	\id\otimes\sigma_z\otimes\sigma_z\otimes\id\otimes\sigma_z\otimes\sigma_z\otimes\sigma_z\otimes\id\otimes\sigma_z\otimes\id\\
&+	\sigma_z\otimes\id\otimes\sigma_z\otimes\sigma_z\otimes\id\otimes\id\otimes\sigma_z\otimes\sigma_z\otimes\id\otimes\sigma_z\\
&+	\sigma_z\otimes\sigma_z\otimes\id\otimes\sigma_z\otimes\sigma_z\otimes\sigma_z\otimes\id\otimes\sigma_z\otimes\sigma_z\otimes\sigma_z\\
&+	\sigma_z\otimes\sigma_z\otimes\sigma_z\otimes\id\otimes\id\otimes\sigma_z\otimes\sigma_z\otimes\id\otimes\id\otimes\sigma_z\\
&+	\sigma_z\otimes\id\otimes\id\otimes\id\otimes\sigma_z\otimes\id\otimes\id\otimes\id\otimes\sigma_z\otimes\sigma_z\\
&+	\id\otimes\sigma_z\otimes\id\otimes\sigma_z\otimes\id\otimes\sigma_z\otimes\id\otimes\sigma_z\otimes\id\otimes\id\\
&+	\id\otimes\id\otimes\sigma_z\otimes\sigma_z\otimes\sigma_z\otimes\id\otimes\sigma_z\otimes\sigma_z\otimes\sigma_z\otimes\id
	\bigl)
	\,.\notag
\end{align}
The second to the fifth summands represent the channels that are used to perfectly win the game.

\begin{figure}
	\centering
	\includegraphics[scale=1]{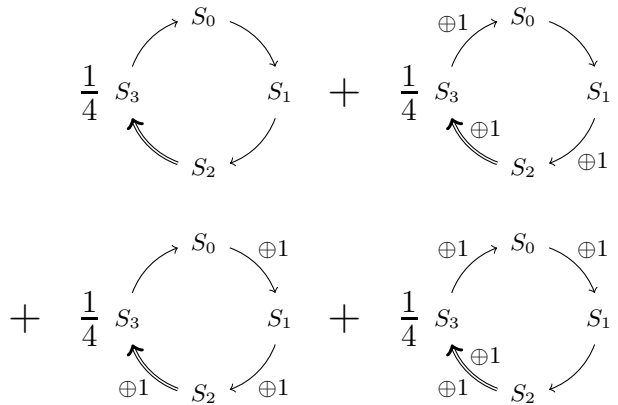}
	\caption{The conditional probability distribution~$W_4$ is a mixture of four circular channels. The channel from~$S_2$ to~$S_3$ is a two-bit channel (double line). If the~$\oplus 1$ operation for the channel from~$S_2$ to~$S_3$ is outside the circle, then the~$\oplus 1$ operation is applied to the first channel, {\em i.e.}, the first bit is flipped, if it is inside the circle, then the~$\oplus 1$ operation is applied to the second channel, {\em i.e.}, the second bit is flipped.}
	\label{fig:w4loops}
\end{figure}

The conditional probability distribution~$W_4$ responds to input~$P_{O_0,O_1,O_2,O_3}(o_0,o_1,o_{2,1},o_{2,2},o_3)=1$ with the following output
\begin{align}
	P_{I_0,I_1,I_2,I_3}&(i_0,i_1,i_2,i_{3,1},i_{3,2})=\\
	&\begin{cases}
		\frac{1}{4},&\parbox[t]{\textwidth}{$i_0=o_3$,~$i_1=o_0$,~$i_2=o_1$,\\$i_{3,1}=o_{2,1}$,~$i_{3,2}=o_{2,2}$,}\\
		\frac{1}{4},&\parbox[t]{\textwidth}{$i_0=\bar o_3$,~$i_1=o_0$,~$i_2=\bar o_1$,\\$i_{3,1}=o_{2,1}$,~$i_{3,2}=\bar o_{2,2}$,}\\
		\frac{1}{4},&\parbox[t]{\textwidth}{$i_0=o_3$,~$i_1=\bar o_0$,~$i_2=\bar o_1$,\\$i_{3,1}=\bar o_{2,1}$,~$i_{3,2}=o_{2,2}$,}\\
		\frac{1}{4},&\parbox[t]{\textwidth}{$i_0=\bar o_3$,~$i_1=\bar o_0$,~$i_2=o_1$,\\$i_{3,1}=\bar o_{2,1}$,~$i_{3,2}=\bar o_{2,2}$,}\\
		0,&\text{otherwise,}
	\end{cases}
\end{align}
where~$o_{2,1},o_{2,2}$ are both bits of the random variable~$O_2$,~$i_{3,1},i_{3,2}$ are both bits of the random variable~$I_3$, and where~$\bar o_i=o_i \oplus 1$.
Therefore,~$W_4$ implements a uniform distribution of four circular channels (see Figure~\ref{fig:w4loops}).

By construction of~$W_4$, no logical paradox arises.
More intuitively, in any strategy that does not break any of the four circular channels of Figure~\ref{fig:w4loops} ({\em i.e.}, every party's output depends on its input), parties~$S_2$ and~$S_3$ use the first bit, the second, or both bits to communicate.
If they use the first bit, then the correlations arising from the first two loops and the last two loops of Figure~\ref{fig:w4loops} interfere constructively.
Both pairs together, however, break the loop.
If they use the second bit, then the first and the third loop, and the second with the last loop yield the same output in every cycle.
In total, all loops cancel each other out.
For the last case as well, where both bits are used for communication, the correlations from the first and the last loop interfere constructively, and so do the second and the third. Ultimately, again, all loops cancel each other out, and no logical paradox can be created.

For larger even~$n$, the conditional probability distribution~$W_n$ as well is constructed out of four loops, as in Figure~\ref{fig:w4loops}, that cancel each other out when one tries to build a logical paradox.
For~$n=2$, the same construction does not work, because the two-bit channel cannot be used to signal from its source to its destination --- it can only be used when combined with other channels. 
In a two-party scenario, however, in order to win the game, each party needs to signal to the other.

\subsection{Winning strategy}
\label{sec:winningstrategy}
For odd~$n$, the strategy~\mbox{$Q^m_i=\hat P_{X_i=x_i,O_i|A_i=a_i,I_i}$} for party~$S_i$ to win the game is
\begin{align}
	Q^m_i&=Q^m_{i,O}\otimes Q^m_{i,I}\\
	&=
	\left(\frac{\id+(-1)^{a'_i}\sigma_z}{2}\right)
	\otimes
	\left(\frac{\id+(-1)^{x_i}\sigma_z}{2}\right)
	\notag
	\,,
\end{align}
where~$a'_i=a_i$ for~$i\equiv m+1 \pmod n$, and~\mbox{$a'_i=a_i+x_i$} otherwise.
The strategies for even~$n$ are equivalent to the strategies for odd~$n$, except that~$S_{n-2}$ has a two-bit output and~$S_{n-1}$ has a two-bit input. Depending on~$M$, they use the first, second, or both bit(s) to receive or send the desired bit.
All local operations are classical since they are diagonal, {\em i.e.}, consist only of measuring and preparing states in the~$\sigma_z$ basis.

The distribution~$P(x_m|a_0,\dots,a_{n-1},M=m)$ is
\begin{align}
	\sum_{\substack{x_i\in\{0,1\}\\i\not=m}}&P(\vec x|\vec a,M=m)\\
	&=\sum_{\substack{x_i\in\{0,1\}\\i\not=m}}\Tr\left(\left(Q^m_{0,I}\otimes\dots\otimes Q^m_{n-1,I}\right.\right.\\
	&\qquad\otimes\left.\left. Q^m_{0,O}\otimes\dots\otimes Q^m_{n-1,O}\right)\cdot W_n\right)\\
	&=\frac{1}{2}\left( 1+(-1)^{X_m+\sum_{i\not=m}A_i} \right)
	\,,
\end{align}
where we rearranged the submatrices of~$Q^m_i$ in the trace expression such that the ordering of the conditional probabilities in~$W_n$ match.
This result is obtained because after taking the trace, each term except~$1$ and~$(-1)^{X_m+\sum_{i\not=m}A_i}$ is either zero or depends on a variable~$X_{i(\not=m)}$ which, in the process of marginalization over~$X_{i(\not=m)}$, cancels out.
For each~$m$, the winning probability is
\begin{align}
	\Pr\left(X_m=\bigoplus_{i\not=m}A_i\,\middle|\,M=m\right)=1
	\notag
	\,.
\end{align}
Therefore, the game is won with certainty.

\subsection{Example:~$n=3$}
The probability of obtaining~$x_0$ in the case~$M=0$ is
\begin{align}
P&\left(x_0|a_0,a_1,a_2,M=0\right)\\
&=\sum_{x_1,x_2\in\{0,1\}}\Tr\left(\left(Q_{0,I}^0\otimes Q_{1,I}^0\otimes Q_{2,I}^0\right.\right.\\
&\qquad\otimes\left.\left. Q_{0,O}^0\otimes Q_{1,O}^0\otimes Q_{2,O}^0\right)\cdot W_3\right)\notag\\
&=
\frac{1}{8}\frac{\Tr\left(\id^{\otimes 6}\right)}{2^6}
\sum_{x_1,x_2\in\{0,1\}}
\left(1+(-1)^{x_0+x_1+x_2+a_0+a_1}\right.\\
&\qquad\left. +\,(-1)^{x_0+a_1+a_2} +(-1)^{x_1+x_2+a_0+a_2}\right)\notag\\
&=\frac{1}{2}\left(1+(-1)^{x_0+a_1+a_2}\right)
\,.
\end{align}
Therefore, the probability of the event~$X_0=A_1\oplus A_2$ is~$1$.
The distribution of~$X_1$ in the case~$M=1$ is
\begin{align}
	P&\left(x_1|a_0,a_1,a_2,M=1\right)\\
	&=\frac{1}{8}\sum_{x_0,x_2\in\{0,1\}}
	\left(1+(-1)^{x_0+x_2+a_0+a_1}\right.\\
	&\qquad\left.+\,(-1)^{x_0+x_1+x_2+a_1+a_2}+(-1)^{x_1+a_0+a_2}\right)\notag\\
	&=\frac{1}{2}\left(1+(-1)^{x_1+a_0+a_2}\right)
	\,,
\end{align}
and, finally, the distribution of~$X_2$ in the case~$M=2$ is
\begin{align}
	P&\left(x_2|a_0,a_1,a_2,M=2\right)\\
	&=\frac{1}{8}\sum_{x_0,x_1\in\{0,1\}}
	\left(1+(-1)^{x_2+a_0+a_1}\right.\\
	&\qquad\left.+\,(-1)^{x_0+x_1+a_1+a_2}+(-1)^{x_0+x_1+x_2+a_0+a_2}\right)\notag\\
	&=\frac{1}{2}\left(1+(-1)^{x_2+a_0+a_1}\right)
	\,.
\end{align}
The probabilities of the events~$X_1=A_0\oplus A_2$ and~\mbox{$X_2=A_0\oplus A_1$} are both~$1$.
Therefore, the game is won with certainty.

Intuitively, in the case~$M=m$, party~$S_{m+1\bmod 3}$ sends~$O_{m+1\bmod 3}=a_{m+1\bmod 3}$ on both circular channels of Figure~\ref{fig:w3loops}.
Thus, party~$S_{m+2\bmod 3}$ receives the uniform mixture of~$I_{m+2\bmod 3}=a_{m+1\bmod 3}$ (left channel of Figure~\ref{fig:w3loops}) and~$I_{m+2\bmod 3}=a_{m+1\bmod 3}\oplus 1$ (right channel of Figure~\ref{fig:w3loops}).
Party~$S_{m+2\bmod 3}$ thereafter sends~$O_{m+2\bmod 3}=I_{m+2\bmod 3}\oplus a_{m+2\bmod 3}$, {\em i.e.},~the uniform mixture of~$O_{m+2\bmod 3}=a_{m+1\bmod 3}\oplus a_{m+2\bmod 3}$ and~$O_{m+2\bmod 3}=a_{m+1\bmod 3}\oplus a_{m+2\bmod 3}\oplus 1$, on both circular channels, yielding the deterministic input~\mbox{$I_m=a_{m+1\bmod 3}\oplus a_{m+2\bmod 3}$} to party~$S_m$.
%Intuitively, in the case~$M=m$, party~$S_{m+1\bmod 3}$ sends~$a_{m+1\bmod 3}$ on both circular channels of Figure~\ref{fig:w3loops}.
%Thus, party~$S_{m+2\bmod 3}$ receives the uniform mixture of~$a_{m+1\bmod 3}$ (left channel of Figure~\ref{fig:w3loops}) and~$a_{m+1\bmod 3}\oplus 1$ (right channel of Figure~\ref{fig:w3loops}).
%Party~$S_{m+2\bmod 3}$ thereafter sends~$I_{m+2\bmod 3}\oplus a_{m+2\bmod 3}$, {\em i.e.},~the uniform mixture of~$a_{m+1\bmod 3}\oplus a_{m+2\bmod 3}$ and~\mbox{$a_{m+1\bmod 3}\oplus a_{m+2\bmod 3}\oplus 1$}, on both circular channels, yielding the deterministic input~\mbox{$I_m=a_{m+1\bmod 3}\oplus a_{m+2\bmod 3}$} to party~$S_m$.

\subsection{Example:~$n=4$}
In the example~$n=4$, we explicitly write the local operations for the third and fourth parties, as the third party has a two-bit output, and the fourth party has a two-bit input.
The local operations for the third party~($S_2$) are
\begin{align}
	Q_2^{m=0}&= \left(\frac{\id\otimes\id+(-1)^{a_2+x_2}\sigma_z\otimes\id}{4}\right) \otimes Q'_2 \notag\\
	Q_2^{m=1}&= \left(\frac{\id\otimes\id+(-1)^{a_2}\sigma_z\otimes\sigma_z}{4}\right) \otimes Q'_2 \notag\\
	Q_2^{m=2}&= \left(\frac{\id\otimes\id}{4}\right) \otimes Q'_2 \notag\\
	Q_2^{m=3}&= \left(\frac{\id\otimes\id+(-1)^{a_2+x_2}\id\otimes\sigma_z}{4}\right) \otimes Q'_2 \notag
	\,,
\end{align}
with
\begin{align}
	Q'_2=\left(\frac{\id+(-1)^{x_2}\sigma_z}{2}\right)
	\,.
\end{align} 
Party~$4$~($S_3$) uses
\begin{align}
	Q_3^{m=0}&= Q'_3 \otimes \left(\frac{\id\otimes\id+(-1)^{x_3}\sigma_z\otimes\id}{2}\right) \notag\\
	Q_3^{m=1}&= Q'_3 \otimes \left(\frac{\id\otimes\id+(-1)^{x_3}\sigma_z\otimes\sigma_z}{2}\right) \notag\\
	Q_3^{m=2}&= \left(\frac{\id+(-1)^{a_3}\sigma_z}{2}\right) \otimes \left(\frac{\id\otimes\id}{2}\right) \notag\\
	Q_3^{m=3}&= Q'_3 \otimes \left(\frac{\id\otimes\id+(-1)^{x_3}\id\otimes\sigma_z}{2}\right) \notag
	\,,
\end{align}
where we use shorthand~$Q'_3$ for
\begin{align}
	Q'_3=\left(\frac{\id+(-1)^{a_3+x_3}\sigma_z}{2}\right)
	\,.
\end{align}

The distributions of~$X_0$,~$X_1$,~$X_2$,~$X_3$, under the condition~$M=0$,~$M=1$,~$M=2$,~$M=3$, respectively, are
\begin{align}
	P(x_0|a_0,a_1,a_2,a_3,M=0)&=\frac{1}{2}\left(1+(-1)^{x_0+a_1+a_2+a_3}\right)\notag\,,\\
	P(x_1|a_0,a_1,a_2,a_3,M=1)&=\frac{1}{2}\left(1+(-1)^{x_1+a_0+a_2+a_3}\right)\notag\,,\\
	P(x_2|a_0,a_1,a_2,a_3,M=2)&=\frac{1}{2}\left(1+(-1)^{x_2+a_0+a_1+a_3}\right)\notag\,,\\
	P(x_3|a_0,a_1,a_2,a_3,M=3)&=\frac{1}{2}\left(1+(-1)^{x_3+a_0+a_1+a_2}\right)\notag
	\,.
\end{align}
Therefore, the event~$X_0=A_1\oplus A_2\oplus A_3$, given~\mbox{$M=0$}, the event~$X_1=A_0\oplus A_2\oplus A_3$, in the case~\mbox{$M=1$}, the event~\mbox{$X_2=A_0\oplus A_1\oplus A_3$}, if~$M=2$, and the event~\mbox{$X_3=A_0\oplus A_1\oplus A_2$} in the case~$M=3$ have probability~$1$.
Which implies that the game is won with certainty.

By consulting Figure~\ref{fig:w4loops}, we can describe the strategy in the following way.
If~$M=m$, then party~$S_{m+1\bmod 4}$ sends~$a_{m+1\bmod 4}$ to the next party by using all four channels of Figure~\ref{fig:w4loops}.
Each of the next two parties in clockwise orientation, {\em i.e.},~party~$S_{m+2\bmod 4}$ and party~$S_{m+3\bmod 4}$, sends the parity of what she receives from the previous party and her input~$(a_{m+2\bmod 4}$,~$a_{m+3\bmod 4}$, respectively).
Depending on~$M$, parties~$S_2$ and~$S_3$ use the first, the second, or both single-bit channels.
In particular, if~$M=0$, then~$S_2$ uses the first channel to communicate to~$S_3$ --- the second channel is ignored.
For~$M=1$ they use both channels, {\em i.e.}, the parity of the inputs to both channels is equal to the bit~$S_2$ sends.
For~$M=2$, the two-bit channel between~$S_2$ and~$S_3$ is ignored.
Finally, for~$M=3$ they use the second channel.
By doing so,~$S_m$ obtains~$a_{m+1\bmod 4}+a_{m+2\bmod 4}+a_{m+3\bmod 4}$, as the introduced bit-flips from the four channels (see Figure~\ref{fig:w4loops}) cancel each other out.

\section{Conclusion}
In an attempt to construct a theory that combines aspects of general relativity and quantum theory, Oreshkov, Costa, and Brukner~\cite{Oreshkov:2012uh} proposed a framework for quantum correlations without causal order.
They proved that some correlations are incompatible with any \emph{a priori\/} causal order and, therefore, are \emph{not compatible with predefined causal order\/} although they satisfy quantum theory {\em locally}.
We consider the classical limit of this framework and show that in sharp contrast to the two-party scenario~\cite{Oreshkov:2012uh}, {\em classical and logically consistent multi-party correlations can be incompatible with any predefined causal order}.
To show this, we propose a game that cannot be won in a scenario with predefined causal order, but is won with certainty when no causal order is fixed.

Recently, the ideas of indefinite causal order and of superpositions of causal orders were applied to quantum computation~\cite{Hardy:2009,Chiribella:2012jg,Colnaghi:2012dv,Chiribella:2013bk,Araujo:2014tb,Morimae:2014ik}.
Furthermore, Aaronson and Watrous~\cite{Aaronson:2009dy} showed that closed timelike curves render classical and quantum computing equivalent. Our result is similar in the sense that the winning probability of the game is the same for the quantum and for the classical framework.
Since the~$W$ object in Figure~\ref{fig:connections} can be thought of as a \emph{channel back in time}, closed timelike curves can be interpreted as being part of the framework.
Closed timelike curves {\em per se\/} are consistent with general relativity~\cite{Godel:1949eb}.
However, Aaronson and Watrous take Deutsch's approach~\cite{Deutsch:1991jo} to closed timelike curves which, as opposed to the framework studied here, is a \emph{non-linear\/} extension of quantum theory --- such extensions are known to allow for communication faster than at the speed of light~\cite{Gisin:1990dh}.

\begin{acknowledgments}
	We thank \v{C}aslav Brukner, Fabio Costa, Christina Giar\-matzi, Issam Ibnouhsein, Ognyan Oreshkov, and Jibran Rashid for helpful discussions.
	We thank \v{C}aslav Brukner for the interpretation of~$W_3$ as two loops.
	We thank Ognyan Oreshkov for pointing out that whenever the matrix~$W$ is diagonal in the computational basis, all local operations can be reduced to objects diagonal in the same basis.
	We thank three anonymous reviewers for helpful comments on the presentation of the results --- especially the last reviewer for the detailed comments.
	The present work was supported by the Swiss National Science Foundation (SNF), the National Centre of Competence in Research ``Quantum Science and Technology'' (QSIT), the COST action on Fundamental Problems in Quantum Physics, the European Commission Project RAQUEL, the John Templeton Foundation, FQXi, and the Austrian Science Fund (FWF) through CoQuS, SFB FoQuS, and the Individual Project 24621.
\end{acknowledgments}

\bibliography{refspra}

\end{document}